\documentclass[aps,pre,reprint,twocolumn,superscriptaddress,showpacs]{revtex4-1}
\usepackage{amssymb,amsmath}
\usepackage[table]{xcolor}
\usepackage{graphicx}
\usepackage{mathtools}
\usepackage[export]{adjustbox}
\usepackage{overpic}
\usepackage[colorlinks=true,
 linkcolor=black,
 urlcolor=blue,
 citecolor=blue]{hyperref}
\usepackage{nicefrac}
\usepackage{multirow}
\usepackage{lipsum} 
\usepackage[normalem]{ulem}
\usepackage{pbox}
\newcolumntype{C}[1]{>{\centering\let\newline\\\arraybackslash\hspace{0pt}}m{#1}}
\usepackage{verbatim}
\usepackage{enumitem}

\usepackage{framed,color}
\definecolor{shadecolor}{rgb}{0.85,0.80,0.80}

\definecolor{myorange}{RGB}{253, 184, 99}
\definecolor{mypurple}{RGB}{178, 171, 210}

\newcommand{\comments}[1]{}
\usepackage{multirow}

\usepackage{float}

\newcommand{\beq}{\begin{equation}}
\newcommand{\eeq}{\end{equation}}
\newcommand{\bal}{\begin{aligned}}
\newcommand{\eal}{\end{aligned}}

\newcommand{\be}{\begin{equation}}
\newcommand{\ee}{\end{equation}}
\newcommand{\bd}{\begin{displaymath}}
\newcommand{\ed}{\end{displaymath}}
\newcommand{\BE}{\begin{eqnarray}}
\newcommand{\EE}{\end{eqnarray}}

\newcommand{\boldpsi}{{\mbox{\boldmath $\psi$}}}

\newcommand{\jb}[1]{\textbf{\textcolor{green}{[JB: #1]}}}

\allowdisplaybreaks
\begin{document}
\title{Consensus, polarisation and coexistence in a continuous opinion dynamics model with quenched disorder}
\author{Joseph W. Baron}
\email{josephbaron@ifisc.uib-scic.es}
\affiliation{Instituto de F{\' i}sica Interdisciplinar y Sistemas Complejos IFISC (CSIC-UIB), 07122 Palma de Mallorca, Spain}

\begin{abstract}
A general model of opinion dynamics is introduced in which each individual's opinion is measured on a bounded continuous spectrum. Each opinion is influenced heterogeneously by every other opinion in the population. It is demonstrated that consensus, polarisation and a spread of moderate opinions are all possible within this model. Using dynamic mean-field theory, we are able to identify the statistical features of the interactions between individuals that give rise to each of the aforementioned emergent phenomena. The nature of the transitions between each of the observed macroscopic states is also studied. It is demonstrated that heterogeneity of interactions between individuals can lead to polarisation, that mostly antagonistic or contrarian interactions can promote consensus at a moderate opinion, and that mostly reinforcing interactions encourage the majority to take an extreme opinion. 
\end{abstract}


\maketitle

\section{Introduction}
Various models of opinion dynamics have been proposed over the years \cite{castellano2009statistical}, each with the aim of providing a microscopic, individual-based mechanism for how a particular macroscopic, population-wide phenomenon can occur. Depending on the macroscopic phenomenon in question (polarisation or consensus of opinion, for example) different mathematical representations of opinion have proved more appropriate or convenient. 

A binary representation of opinion has primarily proved useful for understanding the circumstances under which a population may (or may not) arrive at consensus. Such models are a popular application for the ideas of statistical physics, owing to their similarity with the well-understood Ising model of magnetism \cite{deoliveira2}.  Examples of models that involve a binary representation of opinion include the voter model \cite{clifford, holley1975, vazquez2008}, the Kirman model (or the noisy voter model) \cite{kirman, granovsky} and the majority vote model \cite{deoliveira1, deoliveira2}. 

However, a binary representation of opinion makes it difficult to distinguish between a population with a healthy coexistence of moderate opinions and a more pathologically split population with two camps holding opposing extreme opinions. An exception is made when the two opposing populations have some other distinguishing characteristic, such as belonging to two separate network `cliques' \cite{krueger2017conformity}. Models with many discrete opinions have thus been suggested for the purpose of studying the emergence of polarisation \cite{la2014influence, balenzuela2015undecided, vazquez2020, mobilia2011fixation} and a coexistence of many different opinions \cite{herrerias2019consensus, khalil2021zealots}. 

Models with opinion measured on a continuous spectrum have also been developed. Typically, such models constitute so-called bounded confidence models \cite{castellano2009statistical}, examples of which are the Deffuant \cite{deffuant} and Hegselmann-Krause \cite{hegselmann2002opinion} models. Bounded confidence models were originally conceived for the purpose of modelling how a committee of individuals may come to a consensus or otherwise be grid-locked in two separate opposing camps. Recently, more generalised continuum models with repulsive effects \cite{banisch2019opinion, chen2019modeling, li2017} have been introduced that give rise to a more stark polarisation of opinion.

In this work, it is demonstrated that all of the aforementioned macroscopic phenomena, for which a range of models have previously been developed, can be captured within a single model. Further, the circumstances under which each of these macroscopic phenomena is observed is deduced analytically.

The proposed model represents individual's opinions as varying continuously on a bounded spectrum, with the interactions between individuals being drawn from a fixed distribution. The model allows for individuals to have a mix of antagonistic or reinforcing relationships of various intensities with others. Quenched disorder of this kind is common in models of spin glasses \cite{mezard1987, gardner1985spin, sompolinsky1981dynamic, kirkpatrick1987p, sompolinsky1982relaxational}, complex ecological systems \cite{Galla_2018,  opper1992phase, bunin2017, goel1971volterra} and neural networks \cite{gardner1988space, gardner1988optimal, derrida1987exactly, aljadeff2015transition, sompolinsky1988chaos}, and its effect has also been studied in models of binary opinion dynamics \cite{masuda2010, lafuerza2013, krawiecki2018spin, krawiecki2020ferromagnetic, baron2021persistent}. 

More precisely, we write down dynamical equations for the time-evolution of a population of opinions. Using dynamical mean-field theory \cite{kirkpatrick1987, kirkpatrick1987p, diederich1989replicators, opper1992phase}, we ascertain how the statistics of the interactions between individuals determine the statistics of the resulting population of opinions. Ultimately, we are able to deduce which factors lead to consensus, polarisation and a coexistence of moderate opinions. We are then able to characterise the phase transitions between these macroscopic states. 

In Section \ref{section:model}, we describe the model in detail and provide an interpretation for the model parameters. In Section \ref{section:simulations}, we provide examples of the model leading to consensus, polarisation and a coexistence of moderate opinions. In Section \ref{section:effectiveprocess}, we introduce the dynamic mean-field theory (DMFT) that we use to treat the model analytically. We then use DMFT to deduce the statistics of the population of opinions and the stability of the system in Section \ref{section:fixedpoint}. In Section \ref{section:phasetransitions}, we carefully define polarisation, consensus and coexistence and characterise the phase transitions between these macroscopic states. In Section \ref{section:phasediagrams}, we systematically study the regions of parameter space for which the system exhibits each macroscopic behaviour. Finally, in Section \ref{section:summary} we discuss our findings and conclude. 

\section{Model definition}\label{section:model}
Consider a set of $N$ individuals indexed by $i$. Each individual holds an opinion $x_i$ that varies between two extremes $x_i = 0$ and $x_i = 1$. Each individual's opinion evolves in time according to
\begin{align}
\dot x_i &= g(x_i)\left[\frac{1}{N}\sum_j  x_j - x_i+ \sum_{j\neq i} z_{ij} (x_j -1/2) \right]  , \label{modeleqs}
\end{align}
where $g(x_i)$ is a function which ensures $0\leq x_i \leq 1$ and $z_{ij}$ are quenched Gaussian random variables with the following statistics
\begin{align}
\langle z_{ij} \rangle &= \frac{\mu}{N}, \nonumber \\
\langle (z_{ij}-\mu/N)^2\rangle &= \frac{\sigma^2}{N}, \nonumber \\
\langle (z_{ij} - \mu/N)( z_{ji}-\mu/N) \rangle &= \frac{\Gamma \sigma^2}{N}. \label{statistics}
\end{align}
These moments are scaled with the population size $N$ in such a way as to ensure that the thermodynamic limit $N\to \infty$ can be taken in a sensible fashion \cite{diederich1989replicators, rieger1989solvable, opper1992phase, kirkpatrick1987p, sompolinsky1982relaxational, mezard1987}. The precise form of $g(x_i)$ has little influence on the long-term behaviour of the system as long as $g(x_i) >0$ for $0<x_i<1$, $g(0) = g(1) = 0$ and $g(y + 1/2) = g(1/2 - y)$. This is demonstrated later in Section \ref{section:fixedpoint}. For the sake of the numerical integration, we choose $g(x_i) = x_i(1-x_i)$ (see Appendix \ref{appendix:numerical}).

The opinion dynamics described by Eq.~(\ref{modeleqs}) can be interpreted as follows: The opinion of individual $i$ grows or decays depending on the sign of the term in the square brackets. Individual $i$ is influenced towards the population-averaged opinion $\frac{1}{N} \sum_{j} x_j$, but is also swayed by the opinions of each other individual heterogeneously. The coefficient $z_{ij}$ reflects how individual $i$ is influenced by the opinion of individual $j$. If $i$ holds $j$'s opinion in high regard, then $z_{ij}$ will be positive and $i$ will be influenced to move to join $j$ on the same side of the opinion spectrum (above or below $1/2$). Conversely, if $i$ holds $j$'s opinion in contempt, then $z_{ij}$ will be negative and individual $i$ will be inclined to take a contrary opinion to $j$. 

The parameter $\mu$ in Eq.~(\ref{statistics}) therefore reflects the overall tendency for individuals to be persuaded by others. When $\mu$ is positive, individuals tend to align their opinions with others. When $\mu$ is negative, individuals tend to be more contrarian on average. The system parameter $\sigma$ reflects heterogeneity in the interactions between individuals. When $\sigma$ is large, each individual is likely to have a mix of antagonistic and reinforcing interactions with others. Finally, $0 <\Gamma< 1$ quantifies reciprocity. When $\Gamma$ is close to $1$, if an individual $i$ holds $j$'s opinion in high regard, then $j$ will likely hold $i$ in high regard also. When $\Gamma$ is close to $0$, the influence that $j$ has on $i$ is largely independent of how $i$ affects the opinion of $j$. We do not consider the possibility of $\Gamma<0$ here.

\section{Observed behaviours}\label{section:simulations}
\begin{widetext}
	
	\begin{figure}[H]
		\centering 
		\includegraphics[scale = 0.25]{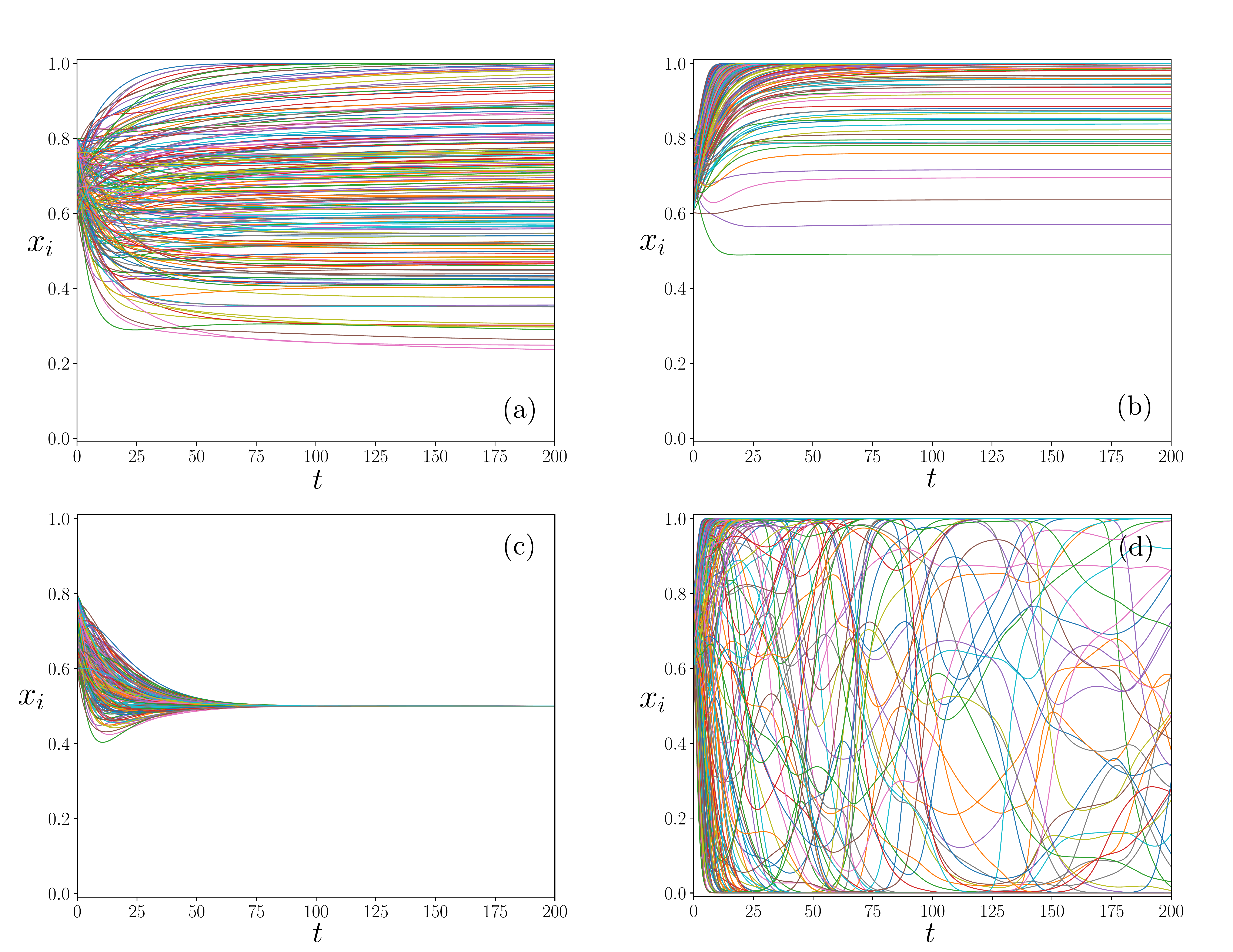}
		\caption{The various behaviours observed by numerically integrating Eqs.~(\ref{modeleqs}) using the method in Appendix \ref{appendix:numerical}. The sets of system parameters for which each of these behaviours are seen are discussed in Sections \ref{section:phasetransitions} and \ref{section:phasediagrams}. One may observe (a) a spread of static opinions with most individuals taking a moderate opinion with $0<x_i<1$ ($\mu = -0.15$, $\sigma = 0.6$, $\Gamma = 0.5$), (b) a majority taking an extreme opinion ($\mu = 0.6$, $\sigma = 0.5$, $\Gamma = 0.5$), (c) consensus at the central opinion ($\mu = -0.5$, $\sigma = 0.6$, $\Gamma = 0.5$), (d) chaotic behaviour where individuals constantly change their opinions over time ($\mu = 0.3$, $\sigma = 2$, $\Gamma = 0.5$). In panel (d), most individuals take either extreme opinion (polarisation), but the majority may take a moderate opinion for certain parameter sets (dynamic coexistence). }\label{fig:simulationresults}
	\end{figure}
\end{widetext}

In Fig. \ref{fig:simulationresults}, the results of integrating Eqs.~(\ref{modeleqs}) numerically are shown for various sets of interaction statistics. In general, we find that the system may exhibit five different behaviours. 

In panel (a) of Fig. \ref{fig:simulationresults}, we observe that individuals' opinions may evolve to reach a fixed point. At the fixed point, the majority of the opinions take a value in the range $0<x_i <1$ in this case. In panel (b), we see that once again the individuals' opinions tend towards a fixed value as $t \to \infty$, but this time most individuals adopt the same extreme opinion. One notes that which extreme this is depends on the initial conditions. In panel (c), we see the possibility for complete consensus at the central opinion $x_i = 1/2$. Finally, in panel (d), we observe that the system may not tend towards a fixed configuration at all, instead exhibiting persistent oscillations in some opinions. In the instance shown in the figure, the population is polarised, with most individuals taking an extreme opinion at any one time. One may also observe similar chaotic behaviour with the majority of individuals adopting a moderate opinion ($0<x_i <1$) at any one time. We refer to this latter state as dynamic coexistence.

In the following sections, we discuss how one can go about deducing the sets of system parameters for which each of these behaviours are exhibited. This is achieved by calculating analytically the average opinion, the spread of opinions and the fraction of individuals holding either extreme opinion.

\section{Dynamic mean-field theory}\label{section:effectiveprocess}
In the thermodynamic limit $N \to \infty$, it is possible to find an ensemble of decoupled single-opinion processes that possess the same statistical properties as the set of coupled processes in Eq.~(\ref{modeleqs}). The existence of such an effective single-opinion process allows one to proceed analytically and deduce the behaviour of the system for various parameter regimes. 

Because the dynamic mean-field theory calculation used to deduce the effective process is lengthy but follows standard methods \cite{Galla_2018, kirkpatrick1987p, diederich1989replicators, opper1992phase}, we only provide an outline of the calculation here and quote the final result. The resulting expression for the effective single-opinion process is verified by comparing with the results of numerically integrating Eqs.~(\ref{modeleqs}) (see Figs. \ref{fig:stationarydistribution},  \ref{fig:transitionchaotic}, \ref{fig:transitionconsensus} and \ref{fig:fraction} ).

One begins by writing down the MSRJD generating functional \cite{altlandsimons, msr, janssen1976lagrangean, dedominicis} for the coupled processes in Eq.~(\ref{modeleqs}) 

\begin{widetext}

\begin{align}
Z\left[ \boldpsi \vert \{z_{ij}\} \right] \propto& \int \mathcal{D}x \mathcal{D}\hat x \, \Omega(x) \exp\left[ i \int dt \psi_i(t) x_i(t) \right] \nonumber \\
&\times\exp\left[ i \int dt \sum_i \hat x_i(t) \left\{ \frac{\dot x_i}{g(x_i)} - \left[ \frac{1}{N}\sum_{j}x_j + \sum_{j}z_{ij}\left(x_j -\frac{1}{2}\right)- x_i \right]  \right\}  \right] , \label{genfunct}
\end{align}
\end{widetext}
where $\{\psi_i(t)\}$ are source variables and $\mathcal{D}x$ and $\mathcal{D}\hat x$ indicate an integral over all possible trajectories of the original opinion variables $\{x_i(t)\}$ and their conjugates $\{\hat x_i(t)\}$ respectively. The pre-factor $\Omega(x)$ is a Jacobian determinant that arises from the fact that paths are constrained to follow Eq.~(\ref{modeleqs}) \cite{brettgalla, altlandsimons}. It does not affect our results, which are valid in the thermodynamic limit $N \to \infty$, so we do not evaluate this pre-factor explicitly here.

Following the standard procedure \cite{Galla_2018, kirkpatrick1987p, diederich1989replicators, opper1992phase}, one takes the average of $Z[\boldpsi \vert \{z_{ij}\} ]$ over all possible combinations of $\{z_{ij}\}$ using the statistics in Eq.~(\ref{statistics}). One then defines appropriate order parameters to decouple the opinions of the $N$ individuals. Finally, one carries out a saddle-point integration to eliminate the `nuisance' auxiliary variables from the generating functional. One then arrives at an expression for the averaged generating functional that corresponds to a set of decoupled stochastic differential equations of the following form 
\begin{widetext}
\begin{align}
\dot x &= g(x) \left\{   M(t) - x + \mu [M(t) - 1/2] + \Gamma  \sigma^2 \int dt' G(t,t')[ x(t') -1/2] + \sigma \eta(t)  \right\} , \nonumber \\
C(t,t') &\equiv \langle \eta(t) \eta(t') \rangle = \langle [x(t)-1/2][ x(t')-1/2] \rangle , \,\,\,\,\,\,\,\, G(t,t') \equiv \left\langle \frac{\delta [x(t)-1/2]}{\delta \eta(t')}\right\rangle , \,\,\,\,\,\,\,\, M(t) \equiv \langle x(t) \rangle ,\label{effectiveprocess}
\end{align}
\end{widetext}
where $\eta(t)$ is a coloured Gaussian noise term and where the angular brackets in the above expressions denote an average over realisations of this noise.

An opinion obeying the effective process in Eq.~(\ref{effectiveprocess}) has identical statistics to any one of the opinions obeying Eqs.~(\ref{modeleqs}) in the limit $N \to \infty$. Thus, if one can deduce the statistics of opinions obeying Eq.~(\ref{effectiveprocess}), one can obtain the statistics of opinions obeying the original model equations Eqs.~(\ref{modeleqs}).

The self-consistent nature of the statistics of the effective single-opinion process in Eq.~(\ref{effectiveprocess}) make it difficult to obtain a full dynamic solution, although one can integrate the effective process numerically \cite{froy}. However, we show in the following section how the statistics of the opinions can be obtained analytically where they each tend towards a fixed value as $t\to\infty$.

\section{Fixed-point solution}\label{section:fixedpoint}
As can be seen from the simulations in Fig. \ref{fig:simulationresults}, there are parameter sets for which all opinions $x_i$ tend towards a stationary value as $t \to \infty$. We show now that the statistics of the opinions at such a fixed point can be obtained from the self-consistent effective process in Eq.~(\ref{effectiveprocess}). 

\subsection{Fixed-point statistics}
Let us begin by defining the fixed-point order parameters
\begin{align}
\chi &= \lim_{t\to\infty}\int^t_0 dt' G(t,t') , \nonumber \\
m &= \lim_{t\to\infty} M(t), \nonumber \\
q &=\lim_{t\to\infty} C(t,t), \nonumber \\
\phi &= \lim_{t\to\infty} \langle \Theta[x(t)]\Theta[1-x(t)] \rangle, \nonumber \\
\phi_0 &= 1 - \lim_{t\to\infty} \langle \Theta[x(t)] \rangle, \nonumber \\
\phi_1 &= 1 - \lim_{t\to\infty} \langle \Theta[1-x(t)] \rangle. \label{statordpar}
\end{align}
where $\Theta[\cdot]$ is the Heaviside function, defined such that $\Theta[0] = 0$.  In addition to defining the fixed-point counterparts to the objects that appear in the effective process Eq.~(\ref{effectiveprocess}), we have also defined the quantities $\phi$, $\phi_0$ and $\phi_1$, which are the fractions of individuals adopting opinions with $0<x_i<1$, $x_i=0$ and $x_i = 1$ respectively.

When the system reaches a fixed point, the noise term $\eta(t)$ in Eq.~(\ref{effectiveprocess}) tends towards a fixed value as $t\to \infty$.  That is, we can write $\lim_{t \to \infty}\eta(t) = \sqrt{q} z$, where $z$ is a standardised Gaussian random variable with $\langle z \rangle = 0$ and $\langle z^2 \rangle =1$.

One observes that there are three solutions that will yield $\dot x = 0$ in Eq.~(\ref{effectiveprocess}): either $x = 0$ or $x = 1$ so that $g(x) = 0$, or the expression in the curly brackets on the right-hand side vanishes. One can show that for each value of the stochastic variable $z$, only one of these solutions is stable (see Appendix \ref{appendix:singlestable}). We thus arrive at the following expression for the unique stable fixed-point value (noting that $z$ is an independent random variable)
\begin{align}
x^\star(z)  &= \begin{cases}
0  \,\,\,&\mathrm{if} \,\,\,\,\,\,\,f(z)\leq 0, \\
f(z)\,\,\, &\mathrm{if} \,\,\,\,\,\,\, 0 < f(z) < 1, \\
1 \,\,\,&\mathrm{if} \,\,\,\,\,\,\, f(z) \geq1 , \label{fpsol1}
\end{cases}
\end{align}
where
\begin{align}
f(z) &=  \frac{1}{2} +\frac{(1 + \mu) (m - 1/2) +  \sigma \sqrt{q} z}{ (1 - \Gamma  \sigma^2 \chi)}.\label{fpsol2}
\end{align}
The order parameters defined in Eq.~(\ref{statordpar}) can thus be determined self-consistently from Eqs.~(\ref{fpsol1}) and (\ref{fpsol2}) by averaging over the stochastic variable $z$. One obtains
\begin{align}
\chi &= \frac{1}{\sigma \sqrt{q}}\left\langle \frac{\partial x^\star}{\partial z}\right\rangle, \nonumber \\
m &= \langle x^\star \rangle, \nonumber \\
q &= \left\langle (x^\star)^2\right\rangle, \nonumber \\
\phi &= \langle \Theta[f(z)]\Theta[1-f(z)] \rangle ,\nonumber \\
\phi_0 &= 1 - \langle \Theta[f(z)] \rangle, \nonumber \\
\phi_1 &= 1 -  \langle \Theta[1-f(z)] \rangle. \label{avz}
\end{align}
Using these expressions, an efficient procedure for finding $m$, $q$, $\chi$, $\phi$, $\phi_0$ and $\phi_1$ as functions of the interaction statistics [$\mu$, $\sigma$ and $\Gamma$ from Eq.~(\ref{statistics})] is described in further detail in Appendix \ref{appendix:solution}. Consequently, we can quantify the fraction of moderates $\phi$, the mean opinion $m$ and the variance of opinions $q - (m-1/2)^2$ in the stationary state. The expressions for these stationary statistics [which are given fully in Eqs.~(\ref{fixedpointsolution}) and (\ref{phis})] are verified in Figs. \ref{fig:transitionchaotic}, \ref{fig:transitionconsensus} and \ref{fig:fraction}.

Let us take the opportunity to emphasise that at no point did we have to use the precise form of $g(x)$ in order to obtain the fixed point solution given in Eqs.~(\ref{fpsol1}) and (\ref{fpsol2}). The only important features of the function $g(x)$ are that $g(0) = g(1) = 0$ and that $g(x) >0$ for $0<x<1$. The precise form of $g(x)$ is immaterial to the calculation of the stationary statistics.

\subsection{Distribution of opinions}

\begin{figure}[b]
	\centering 
	\includegraphics[scale = 0.3]{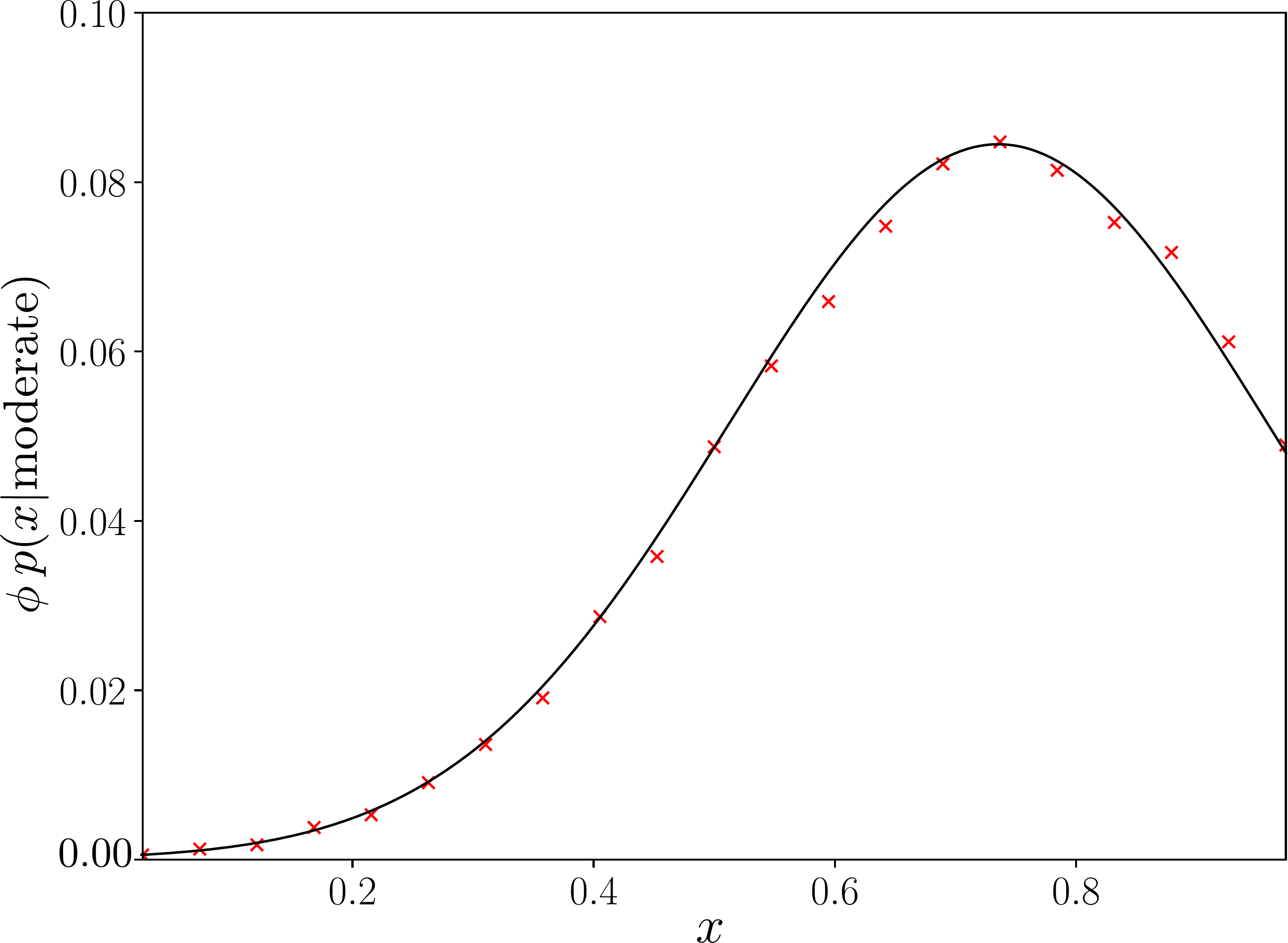}
	\caption{Stationary distribution of opinions when the fixed-point solution is valid. The results of numerically integrating of Eqs.~(\ref{modeleqs}) were averaged over 10 trials to produce the red crosses. The theory prediction in Eq.~(\ref{Gaussiandist}) is shown as a solid black line. The system parameters are $\mu = 0.6$, $\sigma =0.6$, $\Gamma = 0.5$, $N = 200$.  }\label{fig:stationarydistribution}
\end{figure}

From Eqs.~(\ref{fpsol1}) and (\ref{fpsol2}), we see that the stationary opinions are distributed according to a clipped Gaussian distribution. The density of the opinions that do not reside at the extremes $x = 0$ or $x =1$ is therefore given by
\begin{align}
P(x\vert \mathrm{moderate}) = \frac{1}{\phi \sqrt{2 \pi \Sigma^2}}\exp\left[-\frac{-(x-m')^2}{2 \Sigma^2}\right] ,\label{Gaussiandist}
\end{align}
with mode and width (respectively)
\begin{align}
m' &= \frac{1}{2}+ \frac{(1+\mu)(m-1/2)}{(1 -  \Gamma \sigma^2 \chi)} , \nonumber \\
\Sigma^2 &= \frac{\sigma^2 q}{(1 -  \Gamma \sigma^2 \chi)^2}.
\end{align}
This is verified in Fig. \ref{fig:stationarydistribution}.

\section{Phase transitions}\label{section:phasetransitions}
As was described in Section \ref{section:simulations}, there are five distinct macroscopic behaviours possible in the model presented here. Given our analytical characterisation of the fixed point in the previous section, we may now define these behaviours formally and characterise the transitions between them. 

\subsection{Precise definitions of the phases}

We list below each of the possible macroscopic behaviours and define them in terms of the fixed-point order parameters [see Eq.~(\ref{statordpar})]. \\

\textbf{Stable coexistence} -- This occurs when $1/2<\phi<1$ and there is a spread of opinions such that the variance is non-zero, i.e. $q - (m-1/2)^2>0$. See Fig. \ref{fig:simulationresults}a.\\

\textbf{Majority at one extreme} -- In this case, either $\phi_0>1/2$ or $\phi_1>1/2$. This means that the fraction of moderate opinions must be such that $\phi<1/2$. See Fig. \ref{fig:simulationresults}b.\\

\textbf{Consensus at central opinion} -- Here, $\phi = 1$, so no spread of opinions is possible because the distribution of the opinions in the stationary state is Gaussian. Hence, $m = 1/2$ and $q = 1/4$. See Fig. \ref{fig:simulationresults}c.\\

\begin{widetext}
	
	\begin{figure}[H]
		\centering 
		\includegraphics[scale = 0.28]{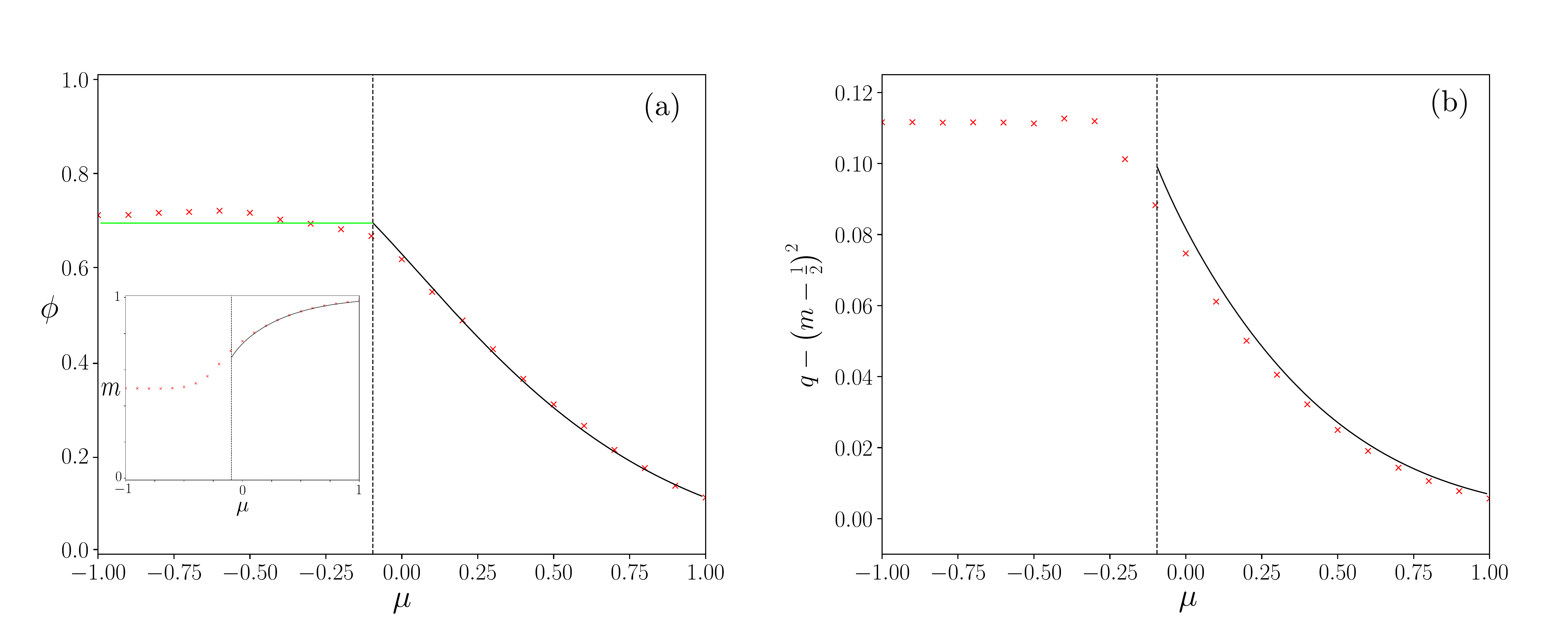}
		\caption{The transition to the dynamic coexistence state (the vertical dashed line depicts the critical point). The transition to polarisation is qualitatively similar, but with $\phi<1/2$ to the left of the vertical line. Red crosses are the results of numerically integrating Eqs.~(\ref{modeleqs}) averaged over 10 trials, the black lines are found by solving Eqs. (\ref{avz}) using the method in Appendix \ref{appendix:solution}, and the horizontal green line in panel (a) is given by Eq.~(\ref{eq:opper}). The system parameters are $\sigma =0.8$, $\Gamma = 0.5$, $N = 400$. Panel (a) shows the conservation of the fraction of moderate opinions $\phi$ despite the instability of the fixed-point solution. The inset shows the mean opinion. Panel (b) shows the dependence of the variance of opinion $q - (m-1/2)^2$ on $\mu$. }\label{fig:transitionchaotic}
	\end{figure}
\end{widetext}

\textbf{Dynamic coexistence} -- In this state, no stable configuration of opinions is reached and the fixed-point solution is invalid. However, the fraction of moderate opinions can still be quantified (see Figs. \ref{fig:transitionchaotic}a and \ref{fig:fraction}). Here, $\phi>1/2$ so that the majority of individuals take a non-extreme opinion at any one time. \\

\textbf{Polarisation} -- Finally, in the polarised state, the fixed-point ansatz is also invalid, but in this case $\phi<1/2$ so that the majority of individuals take an extreme opinion. As is shown in Fig. \ref{fig:simulationresults}d, individuals can switch from one extreme of opinion to the other. As with the dynamic coexistence phase, we can identify precisely when this phase occurs despite the fixed-point solution becoming invalid.

\subsection{Transition to polarisation/dynamic coexistence}\label{subsection:polarisation}
As can be seen from the simulation results in Fig. \ref{fig:simulationresults}d, there are cases where individuals do not settle on a fixed opinion. The point at which the fixed-point solution discussed in Section \ref{section:fixedpoint} becomes unstable and this behaviour emerges can be deduced via linear stability analysis (see Appendix \ref{appendix:LSA}). One arrives at the following compact expression for the set of critical points at which this transition occurs
\begin{align}
\phi = \frac{1}{\sigma^2}\frac{1}{(1+\Gamma)^2} , \label{eq:opper}
\end{align}
where we note that $\phi$ is a non-trivial function of the parameters $\mu$, $\sigma$ and $\Gamma$. 

Holding $\sigma$ and $\Gamma$ constant but varying $\mu$, Fig. \ref{fig:transitionchaotic} demonstrates the nature of the transition into this unstable phase. Imagine that we begin with a large enough value of the agreeableness $\mu$ such that the majority takes one extreme opinion. As we decrease $\mu$ from large positive values, the fraction $\phi$ of individuals adopting moderate opinions increases. At a critical value of $\mu$, we reach a point where Eq.~(\ref{eq:opper}) is satisfied. At this point, the fixed-point solution becomes invalid and the system enters a phase with persistent oscillations in some individual opinions. 

\begin{figure}[H]
	\centering 
	\includegraphics[scale = 0.34]{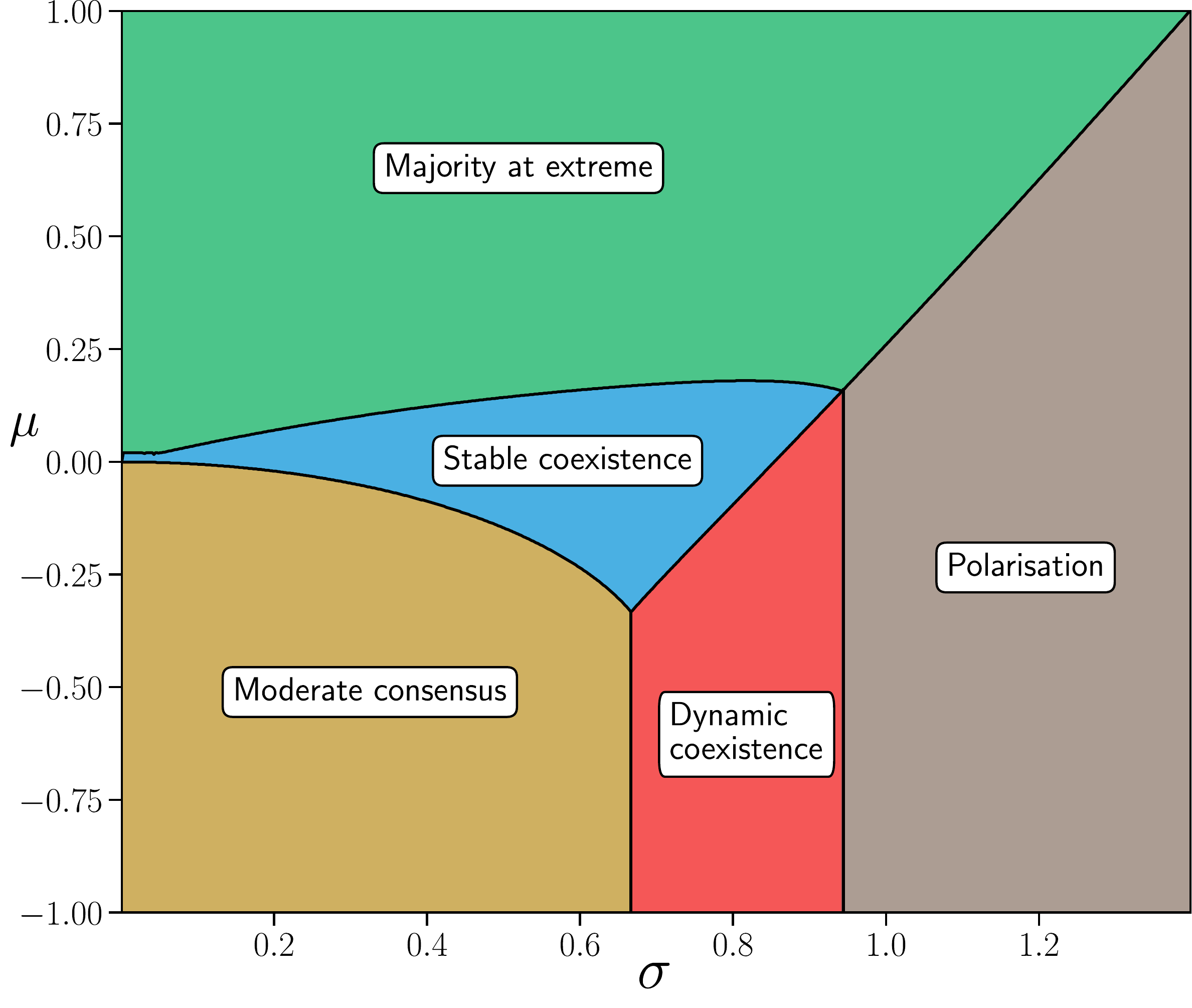}
	\caption{Phase diagram showing the dependence of the emergent behaviour of the population on the statistics of the interactions between individuals. In this case, the five states of the population discussed in Section \ref{section:phasediagrams} are shown as a function of $\mu$ and $\sigma$ for fixed $\Gamma = 0.5$. }\label{fig:muversussigma}
\end{figure}

Depending on the values of $\sigma$ and $\Gamma$ (see Figs. \ref{fig:muversussigma} and \ref{fig:muversusgamma}), this transition may occur when $\phi<1/2$ or when $\phi>1/2$. For $1<\sigma(1+\Gamma)<\sqrt{2}$, the system first transitions into the stable coexistence phase  where $\phi > 1/2$ and then into the dynamic coexistence phase once $\mu$ is reduced beyond the point where Eq.~(\ref{eq:opper}) is satisfied. For $\sigma(1+\Gamma)>\sqrt{2}$, no transition into the stable coexistence phase occurs and instead the population becomes polarised when $\mu$ is decreased beyond the point where Eq.~(\ref{eq:opper}) is satisfied.

We note that although the fixed-point solution becomes invalid when Eq.~(\ref{eq:opper}) is satisfied, the value of $\phi$ is conserved across the transition and into the oscillatory phase, remaining independent of $\mu$ (see Fig. \ref{fig:transitionchaotic} a). The value of $\phi$ in the dynamic coexistence and polarised phases is therefore given by Eq.~(\ref{eq:opper}). This is verified more extensively in Fig. \ref{fig:fraction}. A similar observation has also been noted in the context of ecological models \cite{birolibunin}. This allows us to distinguish between parameter sets for which most individuals adopt an extreme opinion (polarisation, $\phi<1/2$) and those for which most individuals adopt a moderate opinion (dynamic coexistence of opinions, $\phi>1/2$), despite the fixed-point solution being invalid.

\begin{widetext}
	
	\begin{figure}[H]
		\centering 
		\includegraphics[scale = 0.28]{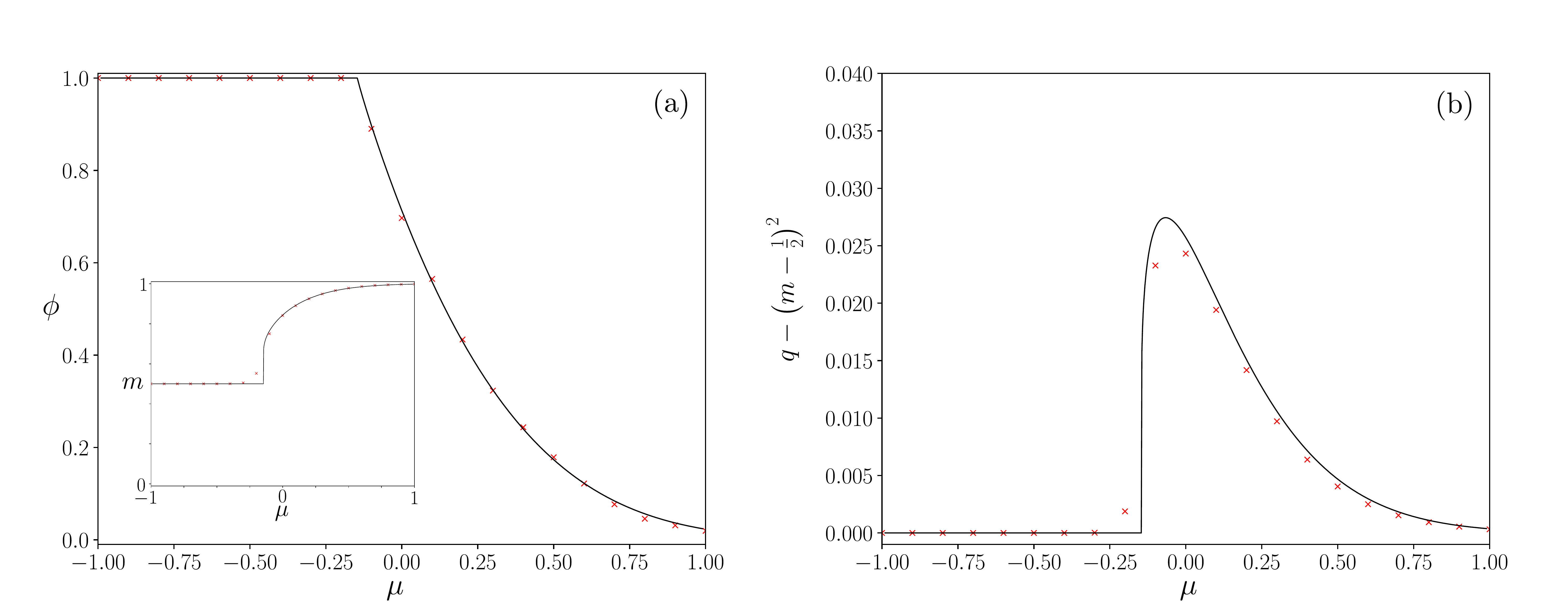}
		\caption{A discontinuous transition to consensus occurs when $\phi =1$. Red crosses are the results of numerically integrating Eqs.~(\ref{modeleqs}) averaged over 10 trials, the black lines were found by solving Eqs. (\ref{avz}) using the method in Appendix \ref{appendix:solution}. The system parameters are $\sigma =0.5$, $\Gamma = 0.5$, $N = 400$. Panel (a) shows the point at which all opinions become moderate ($\phi = 1$) and the transition to consensus occurs. The inset shows how the mean opinion varies across this transition. Panel (b) shows that the variance of opinion $q - (m-1/2)^2$ can be maximised as a function of $\mu$. }\label{fig:transitionconsensus}
	\end{figure}
\end{widetext}

\subsection{Transition to consensus}\label{subsection:consensus}
We saw in the previous subsection that at a certain critical value of $\mu$ the fixed-point solution can become unstable. However, if $\sigma (1 + \Gamma) <1$ [see Eq.~(\ref{eq:opper})], this instability does not occur and $\phi$ is allowed to increase to its maximum value as $\mu$ is reduced (see Fig. \ref{fig:transitionconsensus}a). When $\phi =1$, since the stationary opinions are distributed as a clipped Gaussian, the only possibility is for the trivial solution to Eqs.~(\ref{fixedpointsolution}) to be adopted with $\Delta_0 = 0$ and $\Delta_1 \to \infty$ [see Eqs.~(\ref{deltadef}) in Appendix Section \ref{appendix:solution}]. When this is the case, we have for the mean opinion and variance of opinions respectively $m = 1/2$ and $q - (m-1/2)^2 = 0$. That is, all individuals adopt the central opinion at $x_i = 1/2$ and consensus is achieved (as exemplified in Fig. \ref{fig:simulationresults}c).

So when $\sigma (1 + \Gamma) <1$, the system transitions from the majority at the extreme (at large positive values of $\mu$) to a stable coexistence of opinion (as $\mu$ is reduced). Then, instead of the fixed-point solution becoming unstable as $\mu$ is reduced further, we see a sudden discontinuous jump in the values of the order parameter $m$ (the average opinion) and the variance of opinions $q - (m-1/2)^2$ (see Fig. \ref{fig:transitionconsensus}). This is in contrast to the transition into the dynamic coexistence or polarised phases, where the order parameters change continuously across the transition (Fig. \ref{fig:transitionchaotic}).

\section{From microscopic interactions to macroscopic behaviour}\label{section:phasediagrams} 
Having found a way to quantify the statistics of the opinions when there is a fixed-point solution [see Section \ref{section:fixedpoint}] and having identified the points at which this solution becomes invalid [see Eq.~(\ref{eq:opper})], in Section \ref{section:phasetransitions} we were able to precisely define criteria for each of the behaviours shown in Fig. \ref{fig:simulationresults}. 

Using these criteria, the sets of parameters for which each macroscopic behaviour is observed are explored in more detail in Figs. \ref{fig:muversussigma}, \ref{fig:muversusgamma} and \ref{fig:sigmaversusgamma}. Using the information in these figures, we discuss the role that each of the parameters $\mu$ (agreeableness), $\sigma$ (heterogeneity) and $\Gamma$ (reciprocity) play in governing consensus, polarisation and coexistence.

\subsection{Agreeableness $\mu$}
A large positive degree of agreeableness $\mu$ gives rise to a preponderance of positive values of the coupling constants $z_{ij}$ [see Eq.~(\ref{modeleqs})] and therefore makes individuals inclined to adopt similar opinions. As $\mu$ is increased in comparison to the heterogeneity of interactions $\sigma$, individuals become more extreme (Figs. \ref{fig:transitionchaotic}a and \ref{fig:transitionconsensus}a) and more closely align their views (Figs. \ref{fig:transitionchaotic}b and \ref{fig:transitionconsensus}b). That is, we see the majority at the extreme for large positive $\mu$ and a stable coexistence of opinion for $\mu\approx 0$, as is shown in Figs. \ref{fig:muversussigma} and \ref{fig:muversusgamma}. 

On the other hand, for sufficiently little heterogeneity such that $\sigma(1+\Gamma)<1$, a negative value of $\mu$ fosters consensus at the moderate opinion $x = 1/2$ (Figs. \ref{fig:muversussigma} and \ref{fig:muversusgamma}). Consequently, for fixed values of $\sigma$ and $\Gamma$, there is a value of $\mu$ which maximises the diversity of opinions, as is shown in Fig. \ref{fig:transitionconsensus}b. 

To summarise, if there is sufficiently little heterogeneity so that opinions can form stable arrangements, the sign of $\mu$ dictates where they cluster. For agreeable interactions, the clustering occurs at the extreme. For disagreeable interactions, it occurs at the moderate opinion.

\begin{widetext}
	
	\begin{figure}[h]
		\centering 
		\includegraphics[scale = 0.3]{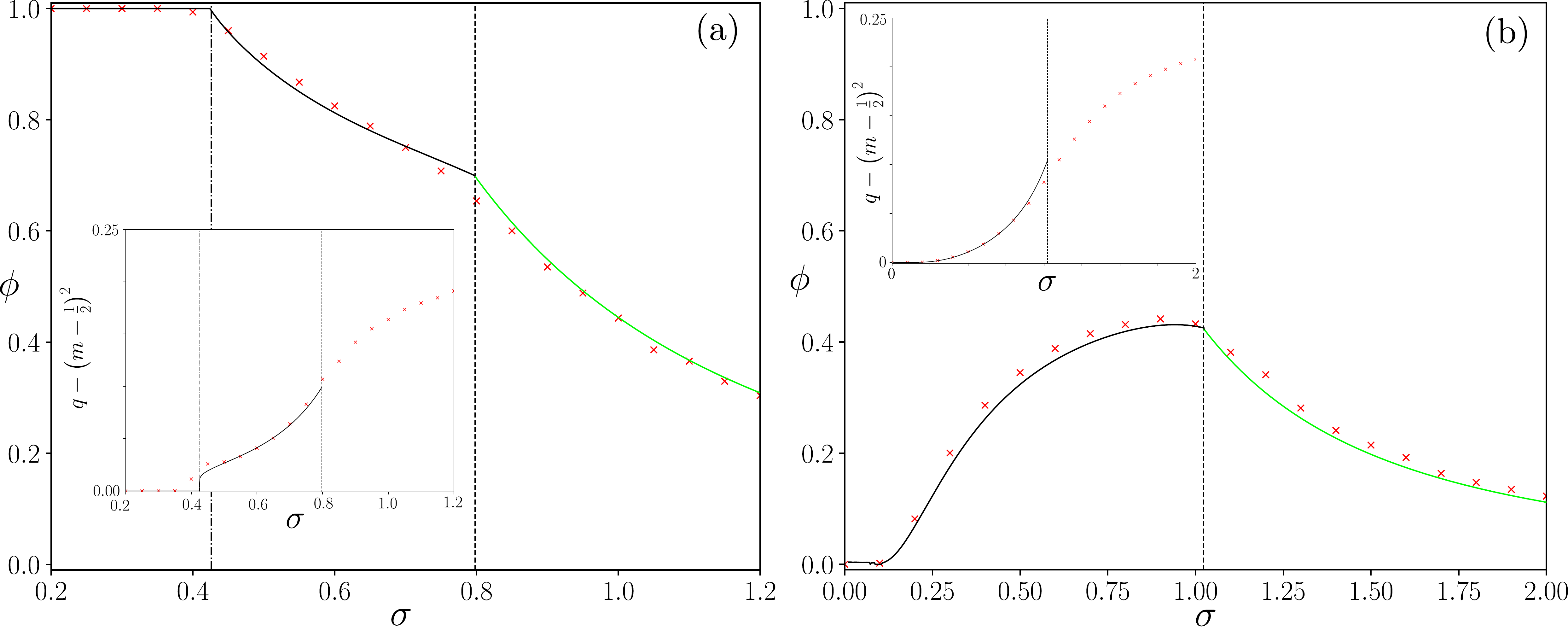}
		\caption{The fraction of moderate opinions $\phi$ versus the heterogeneity $\sigma$. The results of the numerical integration of Eqs.~(\ref{modeleqs}) averaged over 16 trials are shown as red crosses. The black lines are the result of solving Eqs. (\ref{avz}) using the method in Appendix \ref{appendix:solution}. The green lines to the right of the vertical dashed lines are given by Eq.~(\ref{eq:opper}). In panel (a), consensus occurs to the left of the dot-dashed vertical line at $\sigma \approx 0.43$ and dynamic coexistence occurs to the right of the dashed vertical line at $\sigma \approx 0.8$. Polarisation occurs at $\sigma \approx 0.94$ when $\phi$ falls below $1/2$. The remaining system parameters are $N= 400$, $\mu = -0.1$, $\Gamma = 0.5$. In panel (b), there is majority of extreme opinions to the left of the dashed line and polarisation to the right of this line. Here, the remaining system parameters are $N= 400$, $\mu = 0.3$, $\Gamma = 0.5$. The insets in both panels show that the variance (spread of opinion) $q - (m-1/2)^2$ increases as $\sigma$ is increased. }\label{fig:fraction}
	\end{figure}
\end{widetext}

\subsection{Heterogeneity $\sigma$}

Without heterogeneity ($\sigma = 0$), the system would achieve consensus at an extreme opinion for $\mu>0$ or the central opinion for $\mu<0$. As is shown in Fig. \ref{fig:fraction}, increasing $\sigma$ disrupts consensus and increases the spread of opinion. When $\mu>0$, increasing $\sigma$ from zero introduces negative values of $z_{ij}$ (i.e. antagonistic interactions), leading some opinions to stray from the extreme and become more moderate (see Figs. \ref{fig:fraction}b and \ref{fig:sigmaversusgamma}b). When $\mu<0$, increasing $\sigma$ from zero introduces positive values of $z_{ij}$ (i.e. reinforcing interactions), which breaks consensus at $x_i = 1/2$ and again allows for a non-trivial spread of opinion (see Figs. \ref{fig:fraction}a and \ref{fig:sigmaversusgamma}a). 

A large degree of heterogeneity means that each individual has a mixture of repulsive and reinforcing interactions with others. Such a mixture of interaction types has two effects: (1) the formation of two groups is promoted, each with opposing extreme opinions and (2) a kind of frustrated dynamics is fostered where individuals constantly change their opinion. For this reason, heterogeneous interaction promotes the kind of dynamic coexistence or polarisation exemplified in Fig. \ref{fig:simulationresults}d. 

There is always a critical value of $\sigma$ [see Eq. ~(\ref{eq:opper}) and Figs.~\ref{fig:muversussigma} and \ref{fig:sigmaversusgamma}] above which the fixed-point solution becomes invalid and a split in the population begins to occur. As $\sigma$ is increased further beyond this point, the fraction of moderate opinions reduces and the population is influenced towards greater polarisation (as is demonstrated explicitly in Fig. \ref{fig:fraction}). 

Summarising, a greater heterogeneity of interactions causes a greater spread of opinion and is the primary contributing factor to polarisation in the model presented here.

\subsection{Reciprocity $\Gamma$}
A high degree of reciprocity means that the interactions between individuals are likely to be symmetric. This means that individuals will be more influenced to imitate those with whom they have reinforcing links and oppose those with whom they have antagonistic links.

\begin{figure}[H]
	\centering 
	\includegraphics[scale = 0.34]{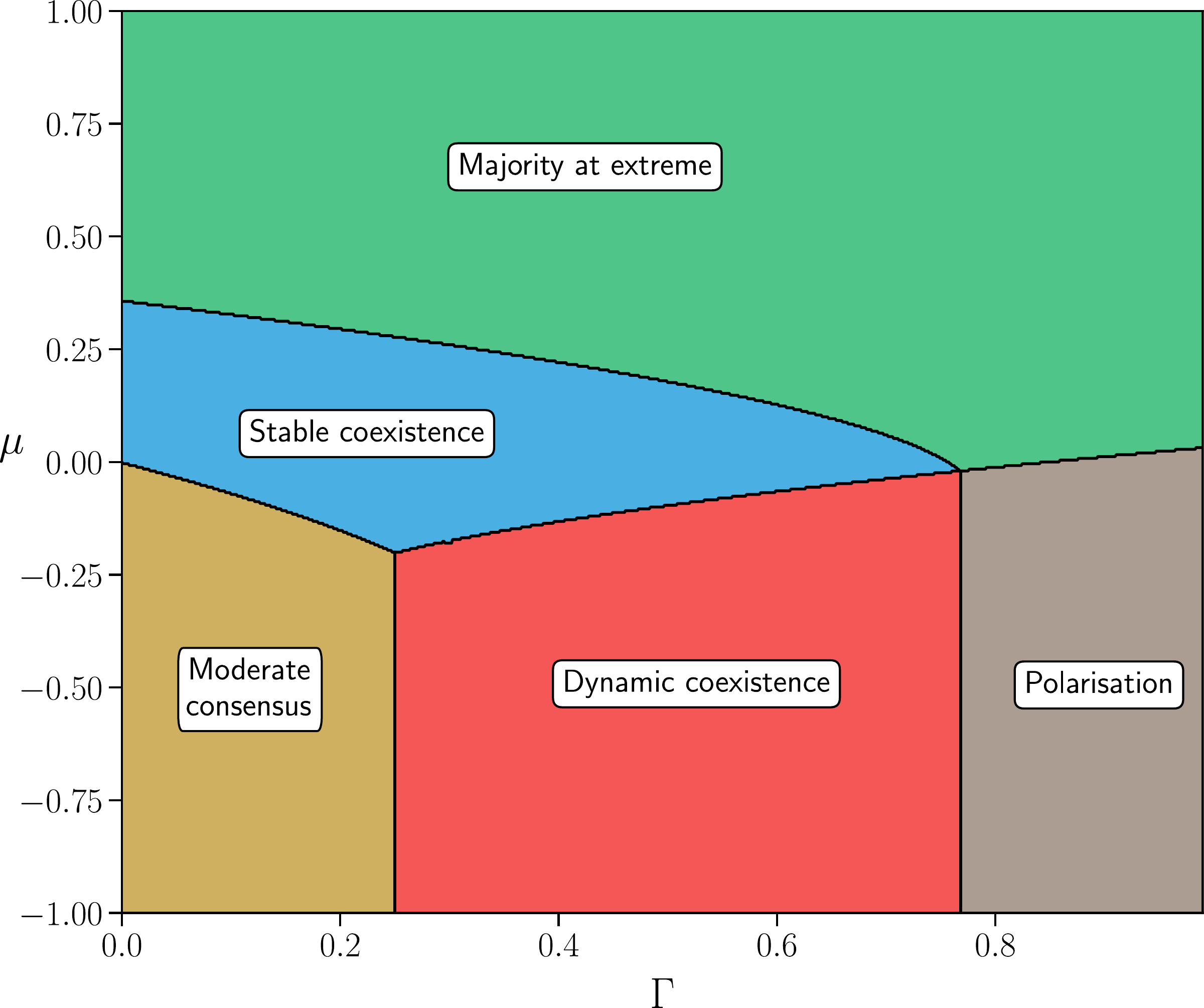}
	\caption{Phase diagram similar to Fig. \ref{fig:muversussigma} showing the dependence on $\Gamma$ instead of $\sigma$. Here $\sigma = 0.8$. The reciprocity $\Gamma$ plays a similar role to the heterogeneity quantified by $\sigma$ (see Fig. \ref{fig:muversussigma}).  }\label{fig:muversusgamma}
\end{figure}

As a result, one effect of increasing the reciprocity $\Gamma$, is the production of a greater number of pairs of both mutually reinforcing opinions and mutually repelling opinions. It is the mixture of both such pairs that leads to the destabilisation of the fixed-point solution and to polarisation. That reciprocity aids heterogeneity in promoting dynamic coexistence and polarisation is demonstrated in Fig. \ref{fig:muversusgamma} and in Eq.~(\ref{eq:opper}). One notes the similar structure of Figs. \ref{fig:muversussigma} and \ref{fig:muversusgamma}.

\begin{widetext}
	
	\begin{figure}[H]
		\centering 
		\includegraphics[scale = 0.32]{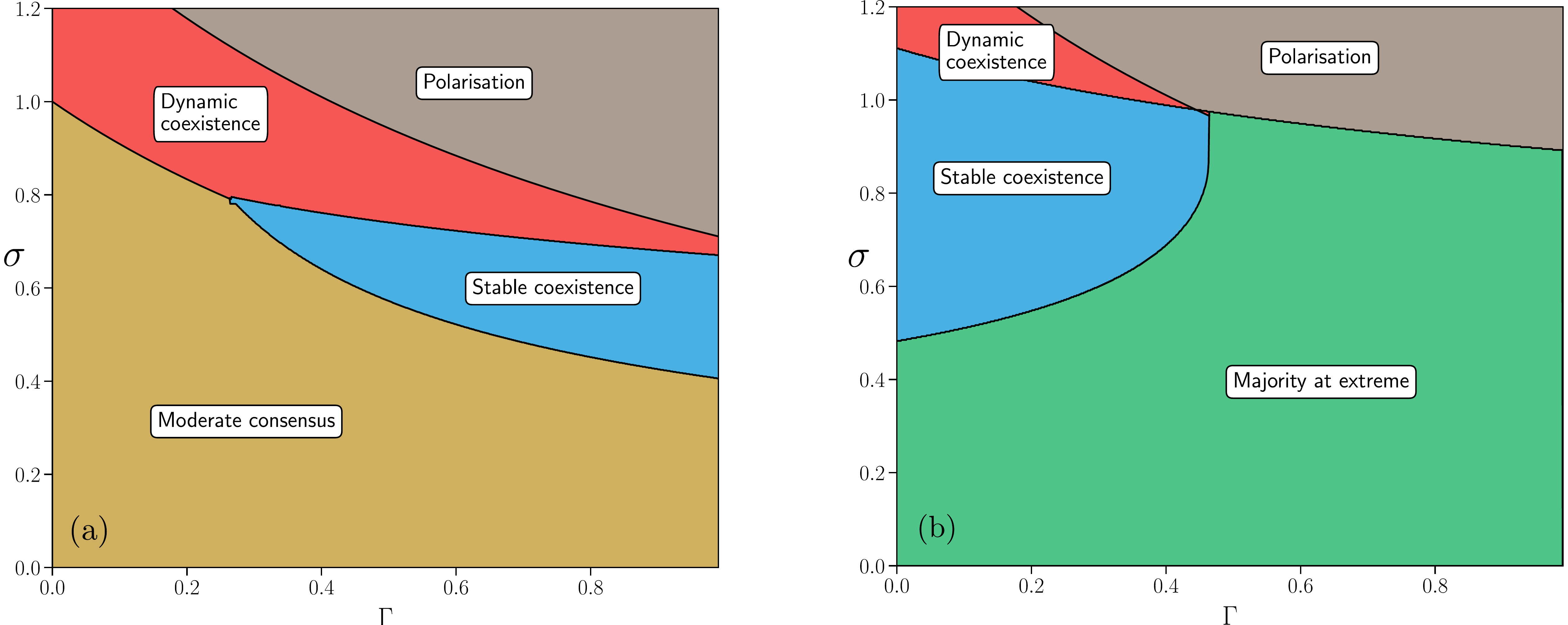}
		
		\caption{Phase diagrams depicting the dependence of the macroscopic behaviour on the microscopic statistics $\sigma$ and $\Gamma$ for fixed $\mu$. It is not possible to have the majority of opinions at the extreme for $\mu<0$ nor moderate consensus for $\mu>0$, so the phase diagrams are qualitatively different for $\mu = -0.2$ (a) and $\mu = 0.2$ (b).   }\label{fig:sigmaversusgamma}
	\end{figure}
	
\end{widetext}

In the region of parameter space where the fixed-point solution is valid, reciprocity has the effect of influencing opinions away from the centre, as is demonstrated in Fig. \ref{fig:sigmaversusgamma}. When $\mu>0$ (Fig. \ref{fig:sigmaversusgamma}b), a greater number of pairs of reinforcing interactions are produced when $\Gamma$ is increased and what was a coexistence of moderate opinions is pushed towards one extreme. When $\mu<0$ (Fig. \ref{fig:sigmaversusgamma}a), increasing $\Gamma$ allows the minority of reinforcing interactions to be paired together, which helps to break consensus at the central opinion. 

In general, reciprocity influences opinions towards the extremes. This extreme could be the same for the entire population when there are mostly reinforcing interactions (so that one finds the majority at one extreme) or it could be both extremes when there is a mix of antagonistic and reinforcing interactions (so that one obtains polarisation).

\section{Discussion and conclusion}\label{section:summary}

In this work, we have introduced a model with continuous opinions and quenched disorder. It was demonstrated that, depending on the precise statistics of the interactions, polarisation, coexistence of opinion (both static and dynamic) and consensus could all be observed. Using dynamic mean field theory, the interaction statistics that led to each of these states were deduced.

The polarisation observed in the model presented here comes about as a direct result of disordered interactions combined with a bounded spectrum for the opinions $x_i$. Because of the bounded opinion interval, when the chaotic transition to polarisation occurs a large proportion of the population may take either one of two extreme opinions, giving rise to a polarised state. This chaotic transition to polarisation (or dynamic coexistence) is analogous to the spin-glass transition \cite{sompolinsky1981dynamic, sompolinsky1982relaxational, kirkpatrick1987p}, to community instability in the context of ecology \cite{bunin2017, birolibunin, diederich1989replicators, galla2006, Galla_2018, opper1992phase} or the transition from quiescence to chaos in neural networks \cite{aljadeff2015transition, sompolinsky1988chaos}. Spin-glass-type transitions have also been identified in binary-state models of opinion dynamics with quenched disorder \cite{krawiecki2018spin, krawiecki2020ferromagnetic}. In all these cases, sufficiently large amounts of disorder in the interactions between individuals (be they magnetic spins, species abundances, neurons or opinions) can lead to a transition to a chaotic or multi-equilibrium state.

Polarisation has previously been shown to be encouraged by a combination of (a) repulsive effects between individuals with differing opinions and (b) reinforcing interactions between individuals with similar opinions. This has been shown in discrete opinion models \cite{la2014influence, vazquez2020, saintier2020model} and in continuous opinion models \cite{banisch2019opinion, chen2019modeling, li2017}. Indeed, a mix of repulsive and reinforcing interactions (marked by high interaction heterogeneity $\sigma$) is what gives rise to polarisation in the present work also. In contrast with these previous works however, we demonstrate that it is not necessary for repulsive/reinforcing interactions to be constrained explicitly to pairs of individuals with opposing/similar opinions respectively. That is, there no need for a dependence of the interactions on the individual opinions themselves, nor is there a need for a confidence interval in order to produce polarisation in the model presented here. 

Aside from polarisation, one effect of the disorder studied here is to disrupt consensus at the extreme opinion by introducing antagonistic interactions, as mentioned in Section \ref{section:phasediagrams}. Antagonistic interactions \cite{baron2021persistent} or the introduction of contrarian agents \cite{khalil2019, banisch2014microscopic, masuda2013voter} also preclude consensus in binary state models. It is a novelty in our model that consensus can be achieved at the central opinion when there is an preponderance of antagonistic interactions (negative $\mu$ and low $\sigma$), a feature that cannot be captured by binary state models by design. We also showed that such consensus at the central opinion can be disrupted by a sufficient amount of interaction heterogeneity.

A stable coexistence of opinion, on the other hand, has previously been shown to be promoted by noise (a random adoption of opinion, independent of social influence) \cite{kirman, herrerias2019consensus} or by imperfect copying \cite{vazquez2019multistate} in discrete opinion models. In this work, we showed that a maximally diverse coexistence of opinion could be achieved (see Figs. \ref{fig:transitionconsensus}, \ref{fig:muversussigma} and \ref{fig:muversusgamma}) by balancing the relative numbers of antagonistic and reinforcing interactions and by ensuring a certain degree of asymmetry in interaction coefficient pairs $(z_{ij}, z_{ji})$. In fact, the diversity of opinions was shown to be increased by greater interaction heterogeneity, with the caveat that too much heterogeneity would destroy stable coexistence and lead to polarisation (see Fig. \ref{fig:fraction}).

Immediate possibilities for future work involve the extension of this model to include more features of real social systems, such as non-trivial social network structure and the possibility of spontaneous changes of opinion. Both of these features have been included in discrete models of opinion dynamics \cite{vazquez2008, suchecki2005, peralta2018, carro2016noisy}. One could also combine the quenched disorder in the interactions that was studied here with heterogeneously biased individuals who harbour a preference for one opinion over another \cite{lafuerza2013, masuda2010}. The interplay between quenched randomness and stochasticity of other origins remains an interesting and general open question.

\begin{acknowledgments}
Partial financial support has been received from the Agencia Estatal de Investigac\' ion (AEI, MCI, Spain) and Fondo Europeo de Desarrollo Regional (FEDER, UE), under Project PACSS (RTI2018-093732-B-C21) and the Maria de Maeztu Program for units of Excellence in R\&D (MDM-2017-0711).
\end{acknowledgments}

\begin{appendix}
\section{Numerical integration}	\label{appendix:numerical}
In order to avoid computational difficulties when any of the opinions $x_i$ approaches the limits of the range $0 \leq x_i \leq 1$ during the integration of Eqs. \ref{modeleqs} [with $g(x_i) = x_i(1-x_i)$], we make the change of variables 
\begin{align}
y_i = \ln\left( \frac{x_i}{1-x_i}\right).
\end{align}
This leads to 
\begin{align}
\dot y_i = \frac{1}{N}\sum_j  \frac{1}{1+e^{-y_j}} - \frac{1}{1+e^{-y_i}}+ \sum_j \frac{ z_{ij}}{2}\frac{1- e^{-y_j}}{1+e^{-y_j}} .
\end{align}
These equations can then be integrated in the usual way using the RK4 method \cite{sulimayers}, avoiding exponential divergences.

\section{Dependence of the fixed-point value on the random variable $z$}\label{appendix:singlestable}
As discussed in Section \ref{section:fixedpoint}, there are three possible stationary solutions to Eq.~(\ref{effectiveprocess}): either $x=0$, $x=1$ or $x= f(z)$, where $f(z)$ is defined in Eq.~(\ref{fpsol2}). We now analyse the stability of each of these three solutions and demonstrate that there is a unique stable fixed-point $x(z)$ corresponding to each value of the stochastic variable $z$, which may take any real value.

Linearising about the possible fixed points and keeping $z$ fixed, we find from Eq.~(\ref{effectiveprocess}) that small deviations about the possible fixed points [$x=0$, $x=1$ and $x = f(z)$ respectively] obey 
\begin{align}
\dot \delta_0 &= \left[1-\chi \phi \sigma^2 \right] f(z) \delta_0, \nonumber \\
\dot \delta_1 &= \left[1-\chi \phi \sigma^2 \right] \left[1-f(z)\right] \delta_1, \nonumber \\
\dot \delta &= g[f(z)]\left[ - \delta + \Gamma \sigma^2 \int dt' G(t,t') \delta(t') + \delta \eta \right] .
\end{align}
From the expression for $\phi(\Delta_0, \Delta_1)$ in Eqs.~(\ref{orderparametersdelta}), we see that we must have $\Delta_1\geq 0$ in order for $\phi$ to be non-negative. It then follows from the definition of $\Delta_1$ in Eq.~(\ref{deltadef}) that $1- \Gamma \sigma^2\chi\geq 0$. The stability of each of the three possible fixed-point values is thus determined by $f(z)$. Let us now explore the three cases $f(z)<0$, $0<f(z)<1$ and $f(z)>1$.

When $f(z)<0$, we have that $\left[1-\chi \phi \sigma^2 \right] f(z)<0$ and $\left[1-\chi \phi \sigma^2 \right] \left[1-f(z)\right]>0$ and $x = f(z)$ is not a valid solution (since $0\leq x \leq 1$). Hence the only valid and stable solution in this case is $x = 0$. 

Similarly, when $f(z)>1$, we have $\left[1-\chi \phi \sigma^2 \right] f(z)>0$ and $\left[1-\chi \phi \sigma^2 \right] \left[1-f(z)\right]<0$  and $x = f(z)$ is again not a valid solution. So in this case the only valid and stable solution is $x = 1$. 

Finally, when $0<f(z)<1$, we have that both $\left[1-\chi \phi \sigma^2 \right] f(z)>0$ and $\left[1-\chi \phi \sigma^2 \right] \left[1-f(z)\right]>0$, so both $x = 0$ and $x = 1$ are unstable solutions. In this case $x = f(z)$ is a valid solution, but there are further conditions that must be met for this solution to be stable (see Appendix \ref{appendix:LSA}). Assuming that these conditions are met, $x = f(z)$ is the only stable solution when $0<f(z)<1$.

In the borderline case $f(z) = 0$, $\left[1-\chi \phi \sigma^2 \right] f(z) =0$ and $\left[1-\chi \phi \sigma^2 \right] \left[1-f(z)\right] >0$, so the only valid and stable solution is $x = f(z) = 0$. A similar argument applies for $f(z) = 1$.

With all the preceding considerations in mind, one finally obtains an expression for the unique stable fixed point as a function of the Gaussian random variable $z$, which is given in Eq.~(\ref{fpsol1}).

\section{Finding the order parameters from the fixed-point solution}\label{appendix:solution}
Beginning with the expression for the fixed points in Eqs.~(\ref{fpsol1}) and (\ref{fpsol2}), we can find the order parameters $\chi$, $m$, $q$, $\phi$, $\phi_0$ and $\phi_1$ self-consistently by averaging over the ensemble of opinions. Using $Dz = dz\, e^{-z^2/2}/\sqrt{2\pi}$ as shorthand for a standardised Gaussian measure, we have from the definitions in Eqs.~(\ref{statordpar})
\begin{widetext}
\begin{align}
\chi &= \frac{1}{\sigma \sqrt{q}} \left\langle \frac{d x^\star}{dz} \right\rangle = \frac{\phi}{1 -  \Gamma \sigma^2\chi} , \nonumber \\
m  &= \langle x^\star \rangle  =\frac{1}{2}+\frac{\sigma \sqrt{q}}{1-\Gamma  \sigma^2 \chi}\int_{\Delta_0-\Delta_1}^{\Delta_0+\Delta_1} Dz (\Delta_0 - z)  + \frac{1}{2} (\phi_1 -  \phi_0 ), \nonumber \\
q &= \langle (x^\star)^2\rangle = \frac{\sigma^2 q }{(1 - \Gamma \sigma^2 \chi)^2} \int^{\Delta_0+\Delta_1}_{\Delta_0 - \Delta_1} Dz (\Delta_0 - z)^2 + \frac{1}{4}(\phi_1 + \phi_0),\label{fixedpointsolution}
\end{align}
\end{widetext}
where we have defined the auxilliary parameters
\begin{align}
\Delta_0 &= (1 + \mu)(m-1/2)/(\sigma \sqrt{q}), \nonumber \\
\Delta_1 &= (1 - \Gamma \sigma^2 \chi)/(2\sigma \sqrt{q}), \label{deltadef}
\end{align}
and we have identified the fractions of individuals holding opinions at the extremes $0$ and $1$ and the fraction of individuals holding moderate opinions (respectively) as
\begin{align}
\phi_0 &= \int_{\Delta_0 + \Delta_1}^{\infty} Dz , \,\,\,\,\,\phi_1 = \int^{\Delta_0 - \Delta_1}_{-\infty} Dz , \nonumber \\
\phi &= \int_{\Delta_0 - \Delta_1}^{\Delta_0+\Delta_1} Dz .\label{phis}
\end{align} 
We wish to solve Eqs.~(\ref{fixedpointsolution}) for the quantities $m$, $\chi$ and $q$ for a given set of interaction statistics [$\mu$, $\sigma$ and $\Gamma$ in Eq.~(\ref{statistics})]. One observes that each of the quantities $m$, $\chi$ and $q$ can be written in terms of only $\Delta_0$ and $\Delta_1$, which are defined in Eq.~(\ref{deltadef}) [see Eq.~(\ref{orderparametersdelta})]. Therefore, following a similar procedure to \cite{Galla_2018}, Eqs.~(\ref{fixedpointsolution}) can be rearranged to yield the following simultaneous equations for $\Delta_0$ and $\Delta_1$ in terms of only the system parameters $\mu$, $\sigma$ and $\Gamma$ 
\begin{align}
\sigma &= (1+\mu)\frac{m(\Delta_0, \Delta_1)-1/2}{\sqrt{q(\Delta_0, \Delta_1)} \Delta_0} , \nonumber \\
\Gamma &= \frac{1-2 \sigma \sqrt{q(\Delta_0, \Delta_1)}\Delta_1}{\sigma^2 \chi(\Delta_0, \Delta_1)},\label{fixedpointsolution2}
\end{align}
where now the order parameters are to be treated as functions of $\Delta_0$ and $\Delta_1$ according to
\begin{widetext}
	
\begin{align}
m(\Delta_0, \Delta_1)  &=\frac{1}{2}+ \frac{1}{2\Delta_1} \left[\frac{1}{\sqrt{2\pi}} \left(e^{- \frac{(\Delta_0 + \Delta_1)^2}{2} } -e^{- \frac{(\Delta_0 - \Delta_1)^2}{2} } \right) + \frac{\Delta_0}{2} \left( \mathrm{Erf}\left[\frac{(\Delta_0 + \Delta_1)}{\sqrt{2}} \right] - \mathrm{Erf}\left[\frac{(\Delta_0 - \Delta_1)}{\sqrt{2}} \right] \right)\right] \nonumber \\
&\,\,\,\,\,\,\,\,\,\,\,\,\,\,\,+ \frac{1}{4} \left\{1 - \mathrm{Erf}\left[\frac{(\Delta_0 + \Delta_1)}{\sqrt{2}} \right] + \mathrm{Erf}\left[\frac{(\Delta_0 - \Delta_1)}{\sqrt{2}} \right] \right\} , \nonumber \\
q(\Delta_0, \Delta_1)  &=  \frac{1}{4 \Delta_1^2} \Bigg[ -\frac{1}{\sqrt{2\pi}}e^{-\frac{(\Delta_0+ \Delta_1)^2}{2}}\left( \Delta_0 (e^{2 \Delta_0 \Delta_1}-1) + \Delta_1  (e^{2 \Delta_0 \Delta_1}+1)\right) \nonumber \\
&\,\,\,\,\,\,\,\,\,\,\,\,\,\,\,\,\,\,\,\,\,\,\,\,\,\,\,\,\,\,+\frac{1 + \Delta_0^2}{2} \left( \mathrm{Erf}\left[\frac{(\Delta_0 + \Delta_1)}{\sqrt{2}} \right] - \mathrm{Erf}\left[\frac{(\Delta_0 - \Delta_1)}{\sqrt{2}} \right] \right)\Bigg]\nonumber \\
&\,\,\,\,\,\,\,\,\,\,\,\,\,\,\,+ \frac{1}{8} \left\{1 + \mathrm{Erf}\left[\frac{(\Delta_0 + \Delta_1)}{\sqrt{2}} \right] + \mathrm{Erf}\left[\frac{(\Delta_0 - \Delta_1)}{\sqrt{2}} \right] \right\}, \nonumber \\
\chi(\Delta_0, \Delta_1) &= \frac{\phi(\Delta_0, \Delta_1)}{2\sigma\sqrt{q(\Delta_0, \Delta_1)}\Delta_1} , \nonumber \\
\phi(\Delta_0, \Delta_1) &= \frac{1}{2}\left\{\mathrm{Erf}\left[\frac{(\Delta_0 + \Delta_1)}{\sqrt{2}} \right] - \mathrm{Erf}\left[\frac{(\Delta_0 - \Delta_1)}{\sqrt{2}} \right] \right\} , \nonumber \\
\phi_0(\Delta_0, \Delta_1) &= \frac{1}{2}\left\{1- \mathrm{Erf}\left[\frac{(\Delta_0 + \Delta_1)}{\sqrt{2}} \right] \right\}, \nonumber \\
\phi_1(\Delta_0, \Delta_1) &= \frac{1}{2}\left\{1+ \mathrm{Erf}\left[\frac{(\Delta_0 - \Delta_1)}{\sqrt{2}} \right] \right\} . \label{orderparametersdelta}
\end{align}
\end{widetext}
Eqs. (\ref{fixedpointsolution2}) can thus be solved numerically for $\Delta_0$ and $\Delta_1$ as a function of the model parameters $\sigma$, $\mu$ and $\Gamma$. All quantities of interest can then be evaluated by plugging the values obtained for $\Delta_0$ and $\Delta_1$ back into Eqs.~(\ref{orderparametersdelta}). By doing so, one can evaluate the order parameters for various values of the model parameters $\sigma$, $\mu$ and $\Gamma$.

\section{Stability of the self-consistent fixed-point solution}\label{appendix:LSA}

Following \cite{Galla_2018, opper1992phase, sompolinsky1982relaxational}, we consider the small deviations about the non-extreme fixed-point solutions $\delta(t) = x(t) - x^\star$ [where $x^\star$ is the fixed-point solution given in Eq.~(\ref{fpsol1})] that arise from the inclusion of a white Gaussian noise term $\xi(t)$ of unit magnitude [that is, $\langle\xi(t) \rangle = 0$ and $\langle \xi(t) \xi(t')\rangle = \delta(t-t')$]. One obtains from Eq.~(\ref{effectiveprocess})
\begin{align}
\dot \delta = x^\star(1-x^\star)\left[ - \delta + \Gamma \sigma^2 \int dt' G(t,t') \delta(t') + \sigma \delta\eta + \xi \right] .
\end{align}
Assuming time-translational invariance of the response function for large $t$ such that $G(t,t') = G(t-t')$ and taking the Fourier transform, one finds
\begin{align}
\tilde \delta = \frac{\sigma\delta\tilde\eta + \tilde \xi}{\frac{i\omega}{x^\star (1-x^\star)}+ [1-\Gamma \sigma^2 \tilde G(\omega)]} .
\end{align}
Finally, squaring both sides and taking the ensemble average, one obtains
\begin{align}
\langle\vert\tilde\delta\vert^2\rangle = \frac{1}{\phi^{-1}\left\vert \frac{i\omega}{x^\star (1-x^\star)}+ [1-\Gamma \sigma^2 \tilde G(\omega)] \right\vert^2 -  \sigma^2 }.
\end{align}
Examining the limit $\omega \to 0$, we see that the denominator diverges (indicating that the fixed-point about which we have linearised becomes invalid) when
\begin{align}
\left(1 -  \Gamma \sigma^2 \chi\right)^2 = \phi \sigma^2 .
\end{align}
Using the first of Eqs.~(\ref{fixedpointsolution}) to eliminate $\phi$, we obtain
\begin{align}
\left( 1 - \Gamma \sigma^2 \chi\right)\left[ 1 - (1 + \Gamma) \sigma^2 \chi\right] = 0
\end{align}
The corresponding value for $\phi$ when $\chi = \frac{1}{\Gamma \sigma^2}$ is $\phi =0$, which does not occur for finite and non-zero values of the system parameters. This means that the instability must occur when [see Eq.~(\ref{eq:opper}) in the main text]
\begin{align}
\chi = \frac{1}{(1+ \Gamma)\sigma^2} \Leftrightarrow \phi = \frac{1}{\sigma^2}\frac{1}{(1+\Gamma)^2} .
\end{align}
\end{appendix}

\end{document}